\documentclass[acmsmall,authorversion]{acmart}

\AtBeginDocument{%
  \providecommand\BibTeX{{%
    \normalfont B\kern-0.5em{\scshape i\kern-0.25em b}\kern-0.8em\TeX}}}
\setcopyright{acmcopyright}
\acmJournal{PACMCGIT}
\acmYear{2021} 
\acmVolume{4} 
\acmNumber{1} 
\acmArticle{} 
\acmMonth{5} 
\acmPrice{15.00}
\acmDOI{10.1145/3451262}

\usepackage[ruled]{algorithm2e}
\usepackage{amsmath}
\usepackage{wrapfig}
\usepackage{xcolor}

\begin{document}

\title{HeterSkinNet: A Heterogeneous Network for Skin Weights Prediction}

\author{Xiaoyu Pan}
\affiliation{%
  \institution{State Key Lab of CAD\&CG, Zhejiang University; ZJU-Tencent Game and Intelligent Graphics Innovation Technology Joint Lab}
  \city{Hangzhou}
  \country{China}}
\email{panxiaoyu6@gmail.com}

\author{Jiancong Huang}
\affiliation{%
  \institution{Tencent Games Lightspeed \& Quantum Studios}
  \city{Shenzhen}
  \country{China}}
\email{bbbhuang@tencent.com}

\author{Jiaming Mai}
\affiliation{%
  \institution{State Key Lab of CAD\&CG, Zhejiang University; ZJU-Tencent Game and Intelligent Graphics Innovation Technology Joint Lab}
  \city{Hangzhou}
  \country{China}}
\email{maijmwq@126.com}

\author{He Wang}
\affiliation{%
  \institution{School of Computing, University of Leeds}
  \city{Leeds}
  \country{United Kindom}}
\email{H.E.Wang@leeds.ac.uk}

\author{Honglin Li}
\affiliation{%
  \institution{Quanzhou Medical College}
  \city{Quanzhou}
  \country{China}}
\email{lihonglin79@qq.com}

\author{Tongkui Su}
\affiliation{%
  \institution{Tencent Games Lightspeed \& Quantum Studios}
  \city{Shenzhen}
  \country{China}}
\email{tongkuisu@tencent.com}

\author{Wenjun Wang}
\affiliation{%
  \institution{Tencent Institute of Games}
  \city{Shenzhen}
  \country{China}}
\email{jamesonwang@tencent.com}

\author{Xiaogang Jin*}
\affiliation{%
  \institution{Corresponding author, State Key Lab of CAD\&CG, Zhejiang University; ZJU-Tencent Game and Intelligent Graphics Innovation Technology Joint Lab}
  \city{HangZhou}
  \country{China}}
\email{jin@cad.zju.edu.cn}

\begin{abstract}
    \pagebreak
  Character rigging is universally needed in computer graphics but notoriously laborious. We present a new method, HeterSkinNet, aiming to fully automate such processes and significantly boost productivity. Given a character mesh and skeleton as input, our method builds a heterogeneous graph that treats the mesh vertices and the skeletal bones as nodes of different types and uses graph convolutions to learn their relationships. To tackle the graph heterogeneity, we propose a new graph network convolution operator that transfers information between heterogeneous nodes. The convolution is based on a new distance \textit{HollowDist} that quantifies the relations between mesh vertices and bones. We show that HeterSkinNet is robust for production characters by providing the ability to incorporate meshes and skeletons with arbitrary topologies and morphologies (e.g., out-of-body bones, disconnected mesh components, etc.). Through exhaustive comparisons, we show that HeterSkinNet outperforms state-of-the-art methods by large margins in terms of rigging accuracy and naturalness. HeterSkinNet provides a solution for effective and robust character rigging.
\end{abstract}



\begin{teaserfigure}
  \includegraphics[width=\textwidth]{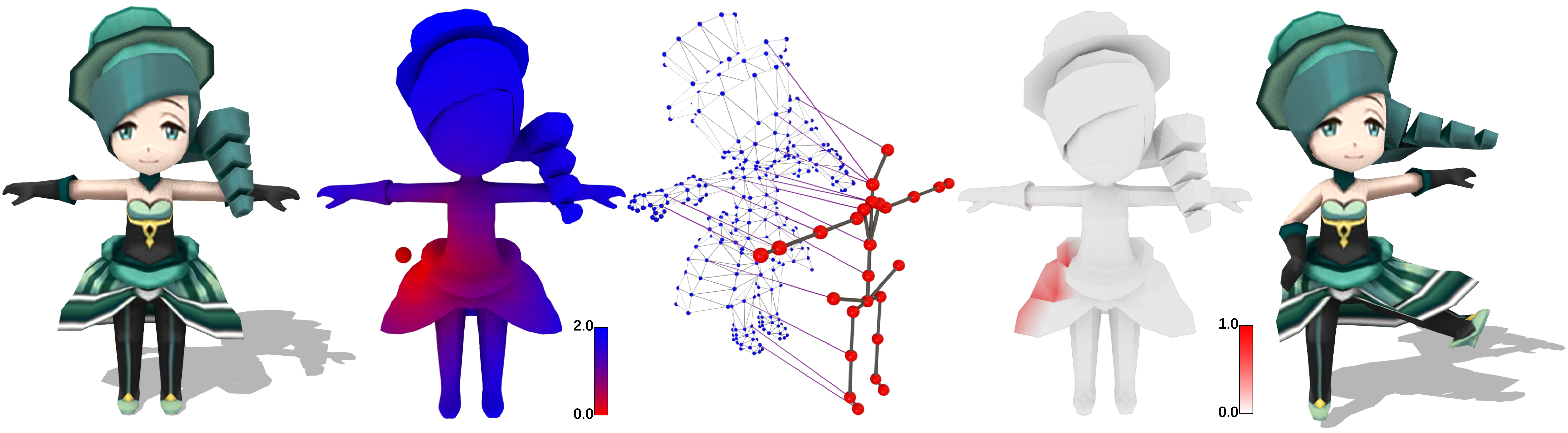}
  \caption{Given a character mesh and skeleton, HeterSkinNet builds a heterogeneous graph network to estimate skin weights. From left to right: example character model, per-vertex HollowDist to the dress bone (red sphere), the heterogeneous graph our network operates on, predicted skin weights of the dress bone, a pose with our estimated skin weights.}
  \label{fig:teaser}
\end{teaserfigure}

\maketitle

\section{Introduction}
There is a surging growth of need for high-quality character animation in film and video game productions. To animate a character mesh, rigging is a popular solution that is widely adopted in industry, where animators first create a skeleton hierarchy and then bind the mesh to the skeleton with the mesh vertices associated with the bones under some weighting schemes, i.e., skin weights. Although methods have been developed to automatically compute skin weights, they are still often painted by hand to enable fine control of the mesh deformation caused by bone transformations. Consequently, when the models and motions become complex, this step becomes time-consuming even for skilled animators.

\begin{figure}[htbp]
  \centering
  \includegraphics[width=\linewidth]{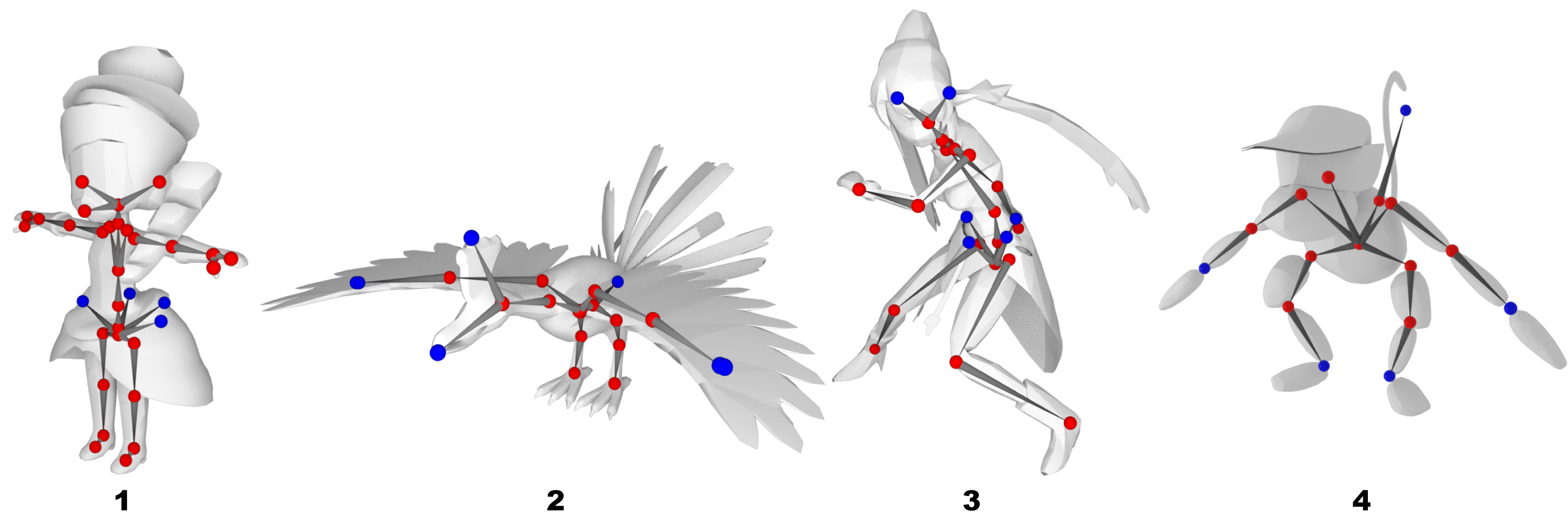}
  \caption{Example characters with their skeletons. The out-of-body bones controlling thin features are colored blue. Most of them are ending bones, which start and end at the same leaf joints.}
  \label{ExampleCharacters}
\end{figure}

Existing efforts have been made to automatically compute the skin weights by hand-crafted functions \cite{baran2007,boneglow,jacobson2011bounded} and learned functions \cite{neuroskinning,xu2020rignet} of vertex features with respect to bones. However, different bones influence areas of various sizes to different extents. While the spine of a character can influence vertices on the torso which are far from it, finger bones only influence close-by vertices. Further, the bone-influence relation changes with the skeleton morphology of characters, shown by Figure \ref{ExampleCharacters} which contains characters of different species with both in-body and out-of-body bones. In extreme cases, the skeleton structure might be totally arbitrary, e.g., out-of-body bones used to deform cloths. The majority of existing methods often simplify the heterogeneous relations between mesh vertices and bones, by homogeneously considering them in one model (e.g., one function or network) to capture different bones influencing different areas, downplaying the importance of the intrinsic shape features of these areas and the connections of the bones. A few attempts \cite{neuroskinning} have been made to apply different functions to different bones by semantically labeling the bones. However, they are designed for one type of skeletal morphology and therefore unable to generalize to characters with different skeletal morphologies.

In this paper, we present a deep-learning-based method, HeterSkinNet, to automatically estimate skin weights of character models with diverse meshes and arbitrary skeletal morphologies. Our network extracts both vertex features and bone features and estimates skin weights according to their relations. It starts by recognizing the different roles of the mesh and the skeleton in rigging, representing them in two graphs, Mesh Graph and Skeleton Graph respectively. Graph convolution operations are used to extract node features. To model the interaction between the mesh and the skeleton, we propose a new intrinsic distance named "HollowDist", which is robust for ill-conditioned situations in production models. Finally, to enable information passing between the two graphs, we propose a new convolution operator, which convolutes between the two kinds of nodes and builds interactions between them. We evaluate HeterSkinNet on a large dataset with a wide variety of meshes and skeletons. The qualitative and quantitative results show that our method can handle the mesh and skeleton diversity well, even for out-of-body bones and disconnected mesh components. We also extensively compare our method with state-of-the-art methods where the quantitative results are improved by large margins.

Formally, our technical contributions include:

\begin{itemize}
  \item The first heterogeneous graph neural network that estimates skin weights for character models with diverse skeletal morphologies and body meshes.
  \item A new graph convolution operator that handles heterogeneous nodes in a graph.
  \item A new distance to quantify vertex-bone relations which can handle ill-conditioned situations including out-of-body bones and disconnected meshes.
\end{itemize}

\section{Related Work}

\paragraph{Skin Deformations.}
Various skinning techniques have been proposed for character rigging. Among them, Linear Blend Skinning (LBS) \cite{LBS} and Dual Quaternion Skinning (DQS) \cite{kavan2007skinning} are widely used in real-time applications due to their simplicity and computational efficiency. In these methods, vertices on the mesh are deformed by weighted transformations of bones. The weights associating vertices with bones determine the quality of deformation. Several attempts have been proposed to automatically estimate high-quality skin weights, which can be categorized into geometry-based and data-driven methods. Geometry-based methods estimate skin weights based on hand-crafted functions of vertex features with respect to bones, such as methods utilizing heat diffusion \cite{baran2007}, illumination model \cite{boneglow}, Laplacian energy \cite{jacobson2011bounded} and volumetric geodesic distance \cite{dionne2013geodesic,dionne2014geodesic}. The hand-crafted functions deployed in these methods make assumptions for the weight distributions. Therefore, it is hard for these methods to capture the anatomic information intrinsic in the input meshes, such as skin flexibility and spine disparity.

Data-driven methods can successfully capture anatomic information of meshes by learning from a set of character models with artist-painted skin weights. NeuroSkinning \cite{neuroskinning} is the first work predicting skin weights of production characters with complex garments. Their network performs graph convolution on graphs represented by meshes to extract high-level features of vertices and predicts the vertex weights as outputs. However, their network assumes that the bones should be labeled semantically as human bones or cloth bones, which restricts its generalization to models with different skeletal morphologies. RigNet \cite{xu2020rignet} provides an end-to-end method of rigging models of arbitrary categories, which predicts the model skeleton and corresponding skin weights jointly. To this end, in the skinning stage of their network, they take intrinsic vertex-bone distances as inputs, which catch the implicit mesh shape information. However, their method does not differentiate the skin weights functions applied for different bones. Different from the two methods mentioned above, our method can be applied to models with arbitrary skeletal morphologies while considering the different situations of bones. Our approach also takes intrinsic vertex-bone distances as network inputs and estimates skin weights using vertex features and bone features extracted by our heterogeneous network.

\paragraph{Graph Neural Networks.}
Graph Neural Networks (GNNs) have shown superior power on deep learning tasks processing data represented by graphs \cite{bronstein2017geometric,zhou2018graph,wu2020comprehensive,zhang2020deep}. There is an increasing interest in applying GNNs on geometric data of different representations \cite{xiao2020survey}, especially for meshes. These GNNs can be generally categorized into spectral approaches \cite{boscaini2016learning,monti2017geometric,yi2017syncspeccnn} and spatial approaches \cite{meshcnn,masci2015geodesic,xu2020rignet,neuroskinning,verma2018feastnet}. Spectral approaches operate on a graph's spectral domain by performing eigen decomposition of the graph Laplacian. They fail to generalize to meshes with different topologies, which have different graph Laplacians. Spatial methods operate on spatially neighboring graph nodes, so that they can generalize to unseen graphs. Our network performs convolution on spatially connected nodes and thus can process character models with different meshes and skeletal morphologies.

In contrast to the above networks operating on homogeneous graphs, other networks focus on handling heterogeneous graphs. Wang et al. \cite{wang2019heterogeneous} proposed a graph neural network based on a hierarchical attention mechanism for general graph analysis tasks. Similarly, Hu et al. \cite{linmei2019heterogeneous} deployed a dual-level attention mechanism on heterogeneous graphs for short text classification. However, our method applies a heterogeneous graph neural network on a different task using different convolution operations, which builds heterogeneous graphs constituted of vertex and bone nodes and learns their high-level features by transferring features with their homogeneous neighbors and heterogeneous neighbors alternatively.

\paragraph{Deep-Learning-based Deformation.}

Several deep learning methods have been proposed for mesh deformation. Some methods process mesh animation sequences. Tan et al. proposed Mesh VAE \cite{tan2017mesh} to extract spatially localized deformation components of mesh deformation, which can be used for shape analysis and synthesis tasks. Qiao et al. introduced \cite{Qiao2018LearningBL} a long-short-term-memory (LSTM) based network to synthesize new mesh animation sequences based on the input sequence.

Some researchers adopt neural networks as non-linear modules to enhance the linear deformations of meshes. Luo et al. \cite{luo2018nnwarp} proposed a network mapping linear displacements to corresponding nonlinear displacements for simulation of elastic materials. Similarly, Bailey et al. \cite{fastBailey} used a fully-connected neural network to approximate the nonlinear deformations of a specific character. Li et al. leveraged a densely connected graph attention network \cite{densegat} for generalization, which can be adopted to human character models with unseen meshes and the same skeleton as those in the training set. Our network estimates the skin weights as the linear portion of deformations, which can be complementary to the above methods.

\begin{figure}[htbp]
  \centering
  \includegraphics[width=\linewidth]{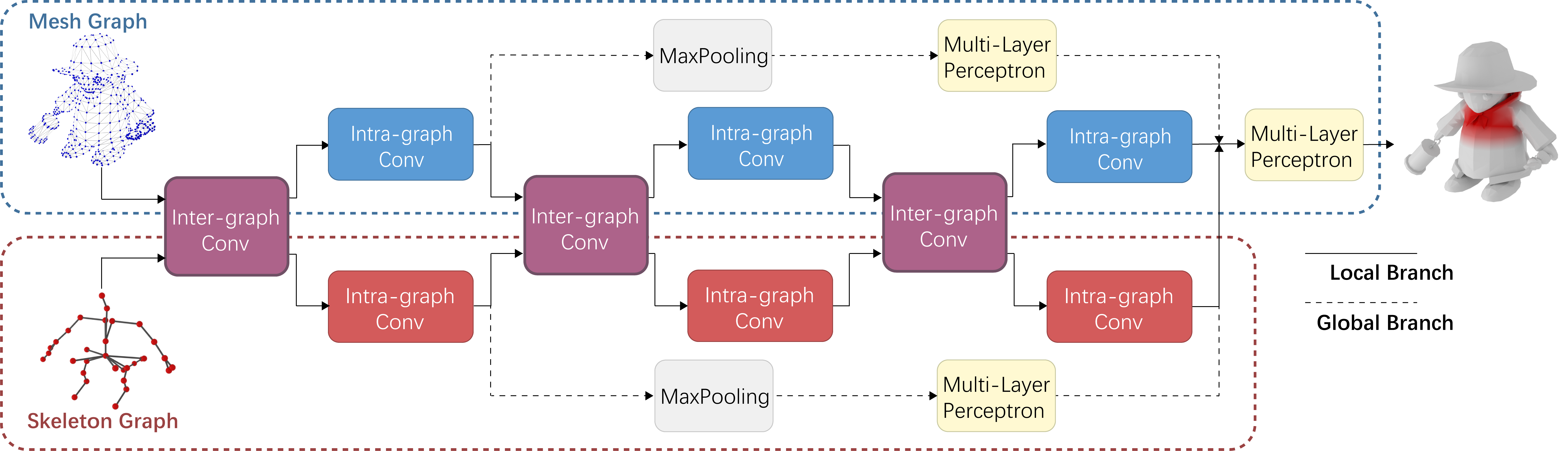}
  \caption{The structure of our network. The network operates on a heterogeneous graph composed of Mesh Graph and Skeleton Graph. Intra-graph convolution components operate within the two subgraphs, and inter-graph convolution components transfer features between them. The detail of inter-graph convolution components is shown in Figure \ref{Inter-graph}.}
  \label{network}
\end{figure}

\pagebreak

\section{Methodology}
Given a character mesh and skeleton, our method aims to obtain the skin weights associating the mesh vertices with the skeleton, which represent the influences of bone transformations on the vertices. First, we present a new distance named `HollowDist`, which quantifies vertex-bone relations. Its calculation is based on surface voxelization and the breadth-first search algorithm. Next, we construct the graph that our network operates on. The graph is constituted of two subgraphs: Mesh Graph, whose nodes and edges represent vertices and mesh edges respectively, and Skeleton Graph, whose nodes and edges indicate bones and joints respectively. The connections between nodes on the two subgraphs are built by HollowDist between corresponding vertices and bones.

Our network is shown in Figure \ref{network} and is composed of intra-graph and inter-graph convolution components: the former operate within subgraphs, i.e., Mesh Graph and Skeleton Graph, and the latter transfer information between the two subgraphs. The network first transfers the node attributes between the two subgraphs via an inter-graph convolution component and then processes the transferred features separately within the two subgraphs by intra-graph components. In each subgraph, the learned features are fed into two branches: a global branch and a local branch. The global branch of each subgraph consists of a max-pooling layer and an MLP layer, which extracts the overall feature of the corresponding subgraph. The local branch is composed of alternatively stacked intra-graph and inter-graph components, which extracts the features of each vertex and bone nodes. At the end of our network, the bone nodes' features are scattered to their neighboring vertex nodes with the concatenation of the two subgraphs' global features, and the resulting features are sent into a stack of MLPs to get the skin weights. The loss function of our network not only minimizes the difference between the predicted skin weights and the ground truth, but also encourages the skin weights to distribute smoothly on the mesh.

In the following subsections, we will first describe the calculation of HollowDist, and then we will introduce the construction of our graph, the individual components of our network and the loss function.

\subsection{HollowDist}

To model vertex-bone relationships, NeuroSkinning \cite{neuroskinning} uses the Euclidean distance which ignores the shape of the mesh and therefore can represent paths that may penetrate the mesh. Thus a restriction needs to be imposed which dictates that the path cannot penetrate the mesh, resulting in a restricted distance. Several distances \cite{earthmover,Rustamov2009InteriorDU} have been proposed in calculating such distances for manifold meshes. But they cannot handle meshes containing non-manifold geometries and intersected triangles. To tackle these meshes, we adopt the idea introduced in \cite{dionne2013geodesic}, which voxelizes the mesh and calculate the distances between bones and vertices via paths inside the mesh. However, their method cannot handle out-of-body bones and meshes with disconnected parts. Different from their method, our approach does not assume that the bones are inside the mesh. We divide the space into mesh cells, bone cells and hollow cells and find paths across hollow cells, which can be either inside the mesh or outside the mesh. Moreover, to tackle disconnected mesh parts, our method restarts the search for unreachable mesh cells.

\begin{wrapfigure}{r}{8cm}
  \centering
  \includegraphics[width=0.6\textwidth]{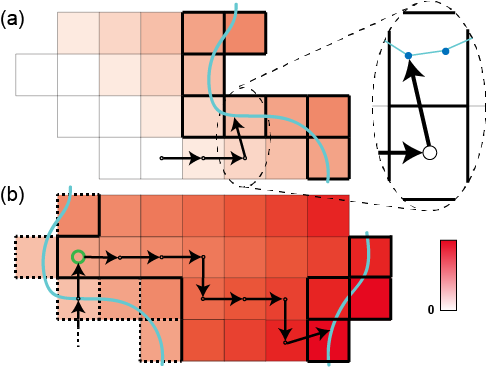}
  \caption{Calculation of the HollowDist. (a) A Path from a bone cell to a mesh cell. The cells are colored to indicate their HollowDist to the bone, in which bone cells are colored white. Mesh surfaces are represented by blue lines, and cells intersecting with it are mesh cells, which have bold boundaries. (b) Restart the search for unconnected mesh cells.}
  \label{HollowDist}
\end{wrapfigure}

We first identify the mesh cells by voxelizing the mesh via the GPU-based surface voxelization method proposed in \cite{schwarz2010fast}, which fills cells intersecting with mesh triangles and keeps the remaining cells hollow. Compared with other rasterization-based methods \cite{voxelization}, this approach does not miss important thin structures in production models, such as garments and hair. Then, we find the path from bone cells to mesh cells via the breadth-first search algorithm, which iteratively traverses from seed cells to their neighboring untraversed cells and marks them as new seeds for the next iteration, extending the path to the new seed cells. Our algorithm is illustrated in Algorithm 1 and is visualized in Figure \ref{HollowDist}. We denote $c$s as cells, with side length of the cell $s_{c}$. For each bone $b_i$, we identify the bone cells as cells intersecting with it. The distances of bone cells are set to zero, and those of the remaining cells are set to infinity (lines 2-9). The path starts from the bone cells and will not penetrate the mesh boundaries which are represented by mesh cells (lines 11-19). For mesh cells that cannot be reached, we find the nearest mesh cell neighboring to any untraversed hollow cell and restart the search from one of its neighboring untraversed hollow cells (lines 20-24). The length of such a path is named HollowDist, as it is mostly on the hollow cells.

The HollowDist for each vertex is calculated based on the length of the path found above. For vertex $v_i$ in cell $c$, its HollowDist to bone $b_j$ is computed as follows:

\begin{equation}
  d_{v_i}^{b_j} = d_{c_{prev}}^{b_j}+||p_{c_{prev}} - p_{v_i}||_2,
\end{equation}
where ${c_{prev}}$ is the cell neighboring to $c$, which is the previous cell on the path, and $p_{c_{prev}}$ is its center position.

\begin{algorithm}[htbp]
  \caption{HollowDist Computation}
  \LinesNumbered
  \KwIn{Cell grid $\mathbb{G}$, side length of cell $s_{c}$ and Character skeleton $\mathcal{S}$}
  \KwOut{Distance between each bone and mesh cells $\textbf{d}$}

  \ForEach{bone $b_i$ in $\mathcal{S}$}{

  \tcp{Initialize cell distances}

  \ForEach{cell $c_i$ of $\mathbb{G}$}{
    $d_{c_i}^{b_i} = \infty$;
  }

  Create empty cell queue $Q$;

  \ForEach{cell $c_i$ intersecting with bone $b_i$}{
    $d_{c_i}^{b_i} = 0$;

    Push $c_i$ to $Q$;
  }

  \tcp{Compute distance}

  \While{exists untraversed mesh cell}{

  \While(){$Q$ is not empty}{

  Pop $c_i$ from $Q$;

  \ForEach(){untraversed cell $c_j$ neighboring to $c_i$}{
  \If(){not ($c_i$ is mesh cell and $c_j$ is not a mesh cell)}{
  $d_{c_j}^{b_i}=d_{c_i}^{b_i}+s_{c}$;

  Push $c_j$ to $Q$;
  }
  }

  }

  \tcp{Restart the search}

  Find the nearest mesh cell $c_i$ neighboring to any untraversed hollow cell;

  \ForEach(){Untraversed hollow cell $c_j$ neighboring to $c_i$}{

  $d_{c_j}^{b_i}=d_{c_i}^{b_i} + s_{c}$;

  Push $c_j$ to $Q$;
  }
  }
  }
\end{algorithm}

\subsection{Graph Construction}

Given a mesh $\mathcal{M}$ with $N$ vertices and its associated skeleton $\mathcal{S}$ with $B$ bones, we construct a heterogeneous graph $\mathcal{G}=(\mathcal{G}_m,\mathcal{G}_s,\mathcal{A}_{ms})$, which is comprised of Mesh Graph $\mathcal{G}_m$, Skeleton Graph $\mathcal{G}_s$ and the adjacency matrix $\mathcal{A}_{ms}$ representing edges between the two subgraphs. The Mesh Graph is denoted as $\mathcal{G}_m=(\mathcal{V}_m,\mathcal{E}_m)$, where $\mathcal{V}_m$ is the set of graph nodes indicating vertices on the mesh, and $\mathcal{E}_m\in \mathcal{V} \times \mathcal{V}$ represents its edges. We also denote $v_i$ as the $i$-th node in $\mathcal{V}_m$. To construct the Skeleton Graph $\mathcal{G}_s=(\mathcal{V}_s,\mathcal{E}_s)$, we first add zero-length helper bones at leaf joints for skinning at the end of limbs, then treat each bone as a node and each joint as an edge, and thus $\mathcal{V}_s$ and $\mathcal{E}_s$ have similar meanings as those in the Mesh Graph $\mathcal{G}_m$. $\mathcal{A}_{ms}\in [0,1]^{N\times B}$ is the connectivity matrix between the Mesh Graph and the Skeleton Graph, $\mathcal{A}_{ms}[i,j]=1$ if vertex $v_i$ is influenced by bone $b_j$. We assume a vertex is influenced by its nearby bones and thus set $\mathcal{A}_{ms}[i,j]=1$ if bone $b_j$ is in the Top-$K$ closest bones to vertex $v_i$ according to HollowDist. $K$ is the max number of influential bones to a vertex, which is set manually.

The inputs of our network are attributes of nodes. The attribute of a vertex node also captures its relationship with the bones. For vertex $v_i$, its calculated HollowDist to all bones are denoted as $\{D_{i,j}\}_{j=1,...,B}$, where $D_{i,j}$ is the HollowDist from $v_i$ to the bone $b_j$. We select the Top-$K$ nearest bones in the ascending order: $\{n_{i,j}\}_{j=1,...,K}$, in which $n_{i,j}$ indicates the index of the $j$-th nearest bone to vertex $v_i$. The vertice's attribute is the concatenation of its position and the inverse of HollowDists to the Top-$K$ closest bones: $f_{v_i}=[p_{v_i}^T, 1/D_{i, n_{i,1}},...,1/D_{i, n_{i,K}}]\in R^{K+3}$. For bone nodes, we use their starting and ending joint positions as attributes: $f_{b_j}=[p_{b_{j,start}}^T, p_{b_{j, end}}^T] \in R^6$.

\subsection{Individual Components}

\paragraph{Intra-graph Convolution Components}

The intra-graph convolution components are built based on EdgeConv \cite{wang2019dynamic}, which aggregates nodes' local features of their homogeneous neighbors. The operations on the Mesh Graph $\mathcal{G}_m$ and the Skeleton Graph $\mathcal{G}_s$ are slightly different. For the Skeleton Graph, we simply use EdgeConv, which is shown as follows:

\begin{equation}
  \begin{split}
    f'_{b_i} & = EdgeConv(b_i,\mathcal{N}(b_i)) \\
   & = \max\limits_{b_j\in{\mathcal{N}(b_i)}} MLP(f_{b_i}||(f_{b_i}-f_{b_j}); W_s).
  \end{split}
  \end{equation}

Here, $||$ is the concatenation operator. $f_{b_i}\in R^{F_b}$ and $f'_{b_i}\in R^{F'_b}$ are the input feature and the output feature of $b_i$, whose dimensions are $F_b$ and $F'_b$, respectively. $\mathcal{N}(b_i)$ is the set of bones connecting to bone $b_i$, and $W_s\in R^{2F_b\times F'_b}$ is the trainable weight matrix of the component.

When applying convolution operations on the mesh graph, a node's receptive field is determined by the distances to its one-ring neighbors, which are influenced by the mesh tessellation. To make the network less sensitive to the mesh tessellation, we adopt the operation introduced in \cite{xu2020rignet}, which separately conducts EdgeConv within nodes' geodesic neighbors and one-ring neighbors and gets the result by processing the concatenation of the results through an MLP. The geodesic neighbors of a vertex are vertices whose geodesic distances from it are within a threshold (0.06). The operation is shown as follows:

\begin{equation}
  \begin{split}
    f'_{v_i} = MLP(\mathop{EdgeConv}(v_i,\mathcal{N}_m(v_i))||\mathop{EdgeConv}(v_i,\mathcal{N}_g(v_i)); W_v),
  \end{split}
  \end{equation}
where $\mathcal{N}_m(v_i)$ and $\mathcal{N}_g(v_i)$ are the one-ring neighbors and geodesic neighbors of the vertex node $v_i$, $W_v\in R^{2F'_v\times F'_v}$ is the trainable weight matrix of the component, and $F_v$ and $F'_v$ denote the dimensions of input features and output features, respectively.

\paragraph{Inter-graph Convolution Component}

\begin{figure}[htbp]
  \centering
  \includegraphics[width=0.85\linewidth]{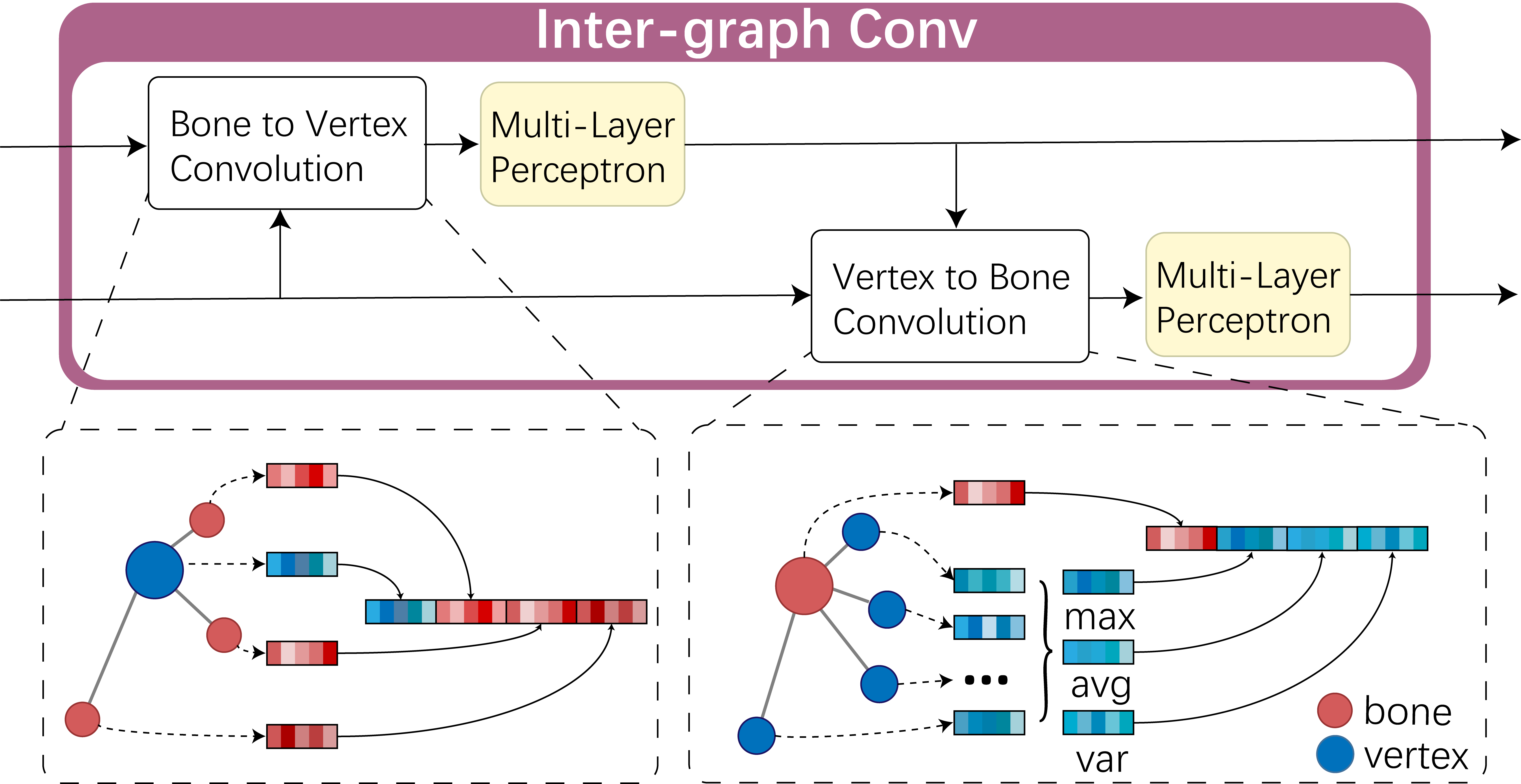}
  \caption{Illustration of the inter-graph convolution operator. Within this operator, the bone-to-vertex convolution and the vertex-to-bone convolution are performed sequentially. In the bone-to-vertex convolution, the vertex nodes in the Mesh Graph concatenate their features with the features of their Top-$K$ nearest bone nodes. In the vertex-to-bone convolution, the bone nodes in the Skeleton Graph concatenate their features with the maximum, mean and variance of their influenced vertex nodes' features. Each convolution is followed by an MLP layer for the convolution operator's nonlinearity.}
  \label{Inter-graph}
\end{figure}

Within this component, the vertex and bone nodes aggregate features from each other mutually. Due to the unbalanced property of $\mathcal{G}$'s two subgraphs, whose numbers of nodes differ greatly, i.e., $\mathcal{G}_m$ contains thousands of vertex nodes while $\mathcal{G}_s$ have dozens of bone nodes, we design the vertex-to-bone and bone-to-vertex convolution operations differently.

For the vertex-to-bone convolution, the bone nodes gather information from their influenced vertex nodes. A lossless way to gather features is to directly concatenate their features. However, the numbers of vertices influenced by bones vary greatly, which makes the dimensions of resulting features indeterminate and thus is unfeasible in a neural network. Inspired by \cite{wen2019pixel2mesh++}, we instead concatenate the maximum, mean and variance pooled from all their influenced vertex nodes' features to make the resulting features' dimensions fixed. This convolution operation also encourages the network to learn information from the correlations among these vertex nodes' features besides learn from the nodes' individual features.

For the bone-to-vertex convolution, the vertex nodes gather features from their corresponding bone nodes. Different from the vertex-to-bone convolution, a vertex is set to be influenced by its nearest $K$ bones, directly concatenating their features can result in fixed-sized features. For a vertex node $v_i$, we first sort the bone nodes according to their HollowDists to it and then concatenate their features with the vertex node's feature and process the concatenated feature via an MLP.

The aforementioned two convolution operations are shown as follows:

\begin{equation}
  f'_{b_j} = MLP((f_{b_j}||\mathop{max}\limits_{v_i\in \mathcal{I}(b_j)}(f_{v_i})||\mathop{mean}\limits_{v_i\in \mathcal{I}(b_j)}(f_{v_i})||\mathop{var}\limits_{v_i\in \mathcal{I}(b_j)}(f_{v_i}));W_{v2b}),
\end{equation}

\begin{equation}
  f'_{v_i} = MLP(f_{v_i}||f_{b_{n_{i,1}}}||...||f_{b_{n_{i,K}}};W_{b2v}).
\end{equation}

Here, $\mathcal{I}(b_j)=\{v_i | \mathcal{A}_{ms}[i,j]=1\}$ represents the set of vertices that bone $b_j$ influences and $W_{v2b} \in R^{F_b\times (F_b+3F_v)}$ and $W_{v2b} \in R^{F_v\times (F_v+K F_b)}$ are learnable weight matrices to make the feature dimensions unchanged after operations.

\subsection{Loss Function}

To make the animation satisfactory, the skin weights should satisfy several constraints. Primarily, they are required to be convex, i.e., $w_{ij}\ge 0$ and $\sum_j w_{ij}=1$, where $w_{ij}$ indicates the skin weight of $v_i$ regarding to bone $b_j$. Rather than directly imposing the constraints on the resulting weight matrix, we add a softmax layer at the end of our network to ensure its non-negativity and affinity. Apart from being convex, the weight matrix is required to be sparse to ensure computation efficiency. In our network, this constraint is automatically satisfied by only predicting the weights regarding to the nearest $K$ bones for each vertex.

\pagebreak

Our loss function is shown as follows:

\begin{equation}
  \mathcal{L} = \mathcal{L}_D + \lambda_S \mathcal{L}_S,
\end{equation}
where $\mathcal{L}_D$ is the data fitting term, $\mathcal{L}_S$ is the smoothing term, and $\lambda_S$ is a predefined smooth factor.

The data fitting term $\mathcal{L}_D$ encourages the predicted weights to be close to the ground truth. We adopt the strategy in \cite{neuroskinning}, which treats the skin weights as label distributions. For a vertex $v_i$, its predicted weight $w_{ij}$ can be considered as its possibility for selecting bone $b_j$. Therefore, our data fitting object is minimizing the distance between the predicted weight distribution and the ground truth distribution. To this end, we use the Kullback-Leibler divergence between the two distributions, $\mathcal{L}_D=\sum w_{ij} log(\frac{w_{ij}}{\hat{w}_{ij}})$, as the data fitting term, where $\hat{w}_{ij}$ is the ground truth.

To encourage the skin weights to distribute smoothly over the mesh, we add a smoothing term $\mathcal{L}_S$ to the loss function, which is calculated using the discrete Laplacian matrix of the mesh: $\mathcal{L}_S=\sum_{j=1}^{B} w^T_j L w_j$, in which $L \in R^{N\times N}$ is the mesh Laplacian matrix, and $w_j\in R^N$ is the weight column vector of bone $b_j$.

\section{Experiments}
\subsection{Dataset}

We use the "ModelsResource-RigNetv1" dataset from \cite{xu2019predicting} for our experiments. This dataset contains manually rigged models with various meshes and different skeletons, covering a wide range of classes, such as humanoids, birds, fish, etc. Most characters' meshes are non-manifold and contain disjoint parts, and some of their bones are placed outside of the meshes. The models are all oriented consistently and rescaled to $1$ meter in height. Some models with multiple parts may have duplicate vertices in the same position at the connection of two parts, which have the same skin weights but different connections. To ensure that they deform in the same way, we merge the vertices in the same position when building the mesh graph and assign the same prediction results to the vertices in that position. Moreover, to ensure that the models with different numbers of vertices have similar impacts on our network during training, for models with less than 1K vertices, we subdivide their meshes so that their vertex numbers range from 1K to 5K. All the above pre-processing steps are performed via Houdini \cite{houdini}.  Finally, 3026 models are randomly selected for training, and the remaining 100 models are used for testing. We present 4 character models in our test set in Figure \ref{ExampleCharacters}, whose statistics are shown in Table \ref{statistics_example}.

\begin{table}[htbp]
  \caption{The statistics of our example characters (Figure \ref{ExampleCharacters})}
  \begin{tabular}{c|cccc}
    \toprule
    Character        & Vertices  & Triangles & Bones & Parts \\
    \midrule
    1          & 4520     & 4455  & 34   &  7 \\
    2          & 4421     & 4341  & 23    &  15 \\
    3          & 3324     & 3084  & 25    & 13 \\
    4          & 3674     & 3564  & 18    &  17 \\
    \bottomrule
  \end{tabular}
  \label{statistics_example}
\end{table}

\subsection{Implementation Details}

For the calculation of HollowDist, we voxelize all meshes using grid size $88\times 88\times 88$. We implement HeterSkinNet using Pytorch Geometric \cite{Fey/Lenssen/2019}, and train it on an NVIDIA GeForce RTX 2080Ti GPU, using Adam optimizer \cite{adam} with learning rate $1e-4$ and weight decay $1e-4$. The dimensions of local and global features of vertex nodes are 256 and 512, and those of bone nodes' features are 128 and 512, respectively. The final layer of the network is a stack of 3 fully connected layers with feature dimensions of $[1024, 512, K]$. In the inter-graph convolution component, the variances of vertex nodes' features gathered by bone nodes are multiplied by $0.1$ for training stability. We use Leaky ReLU with $\alpha = 0.2$ as the activation function in the linear layers of our network. Before training, they are initialized using Kaiming normal \cite{he2015delving}. During training, we randomly drop out 85\% edges in the Mesh Graph and 50\% parameters in the final layers to enhance the generalization ability of our network and save GPU memory. 

\begin{wrapfigure}{r}{7.7cm}
  \centering
  \includegraphics[width=0.5\textwidth]{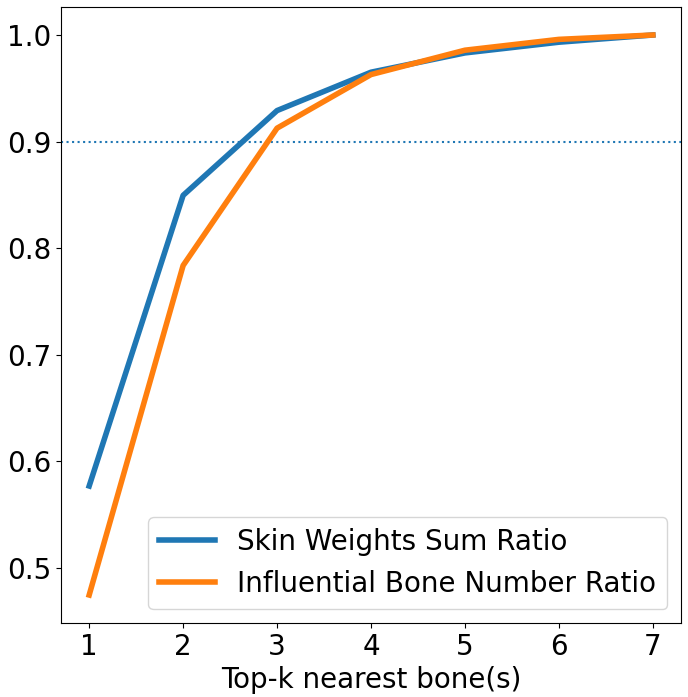}
  \caption{For models in the ModelsResource-RigNetv1 dataset \cite{xu2019predicting}, we calculate the statistical ratio of the skin weights sum of Top-$K$ nearest bones to that of all bones, and the statistical ratio of influential bones in the Top-$K$ nearest bones to all influential bones.}
  \label{top-k}
\end{wrapfigure}

For models in the ModelsResource-RigNetv1 dataset \cite{xu2019predicting}, we calculate the statistical ratio of vertices' skin weights sum of their Top-$K$ nearest bones to that of all bones (the blue line in Figure \ref{top-k}) as $\sum_{i=1}^K w_{i}/\sum w_i$, where $w_i$ is a vertex's skin weight to its $i$-th nearest bone, and the statistical ratio of influential bones among vertices' Top-$K$ nearest bones to that of all influential bones (the orange line in Figure \ref{top-k}) as $=|\mathcal{I}_K|/|\mathcal{I}|$, where $\mathcal{I}_K$ is the set of influential bones in the Top-$K$ nearest bones $\mathcal{I}_K=\{b_i|i<=K\ \text{and}\ w_i>10^{-4}\}$ and $\mathcal{I}$ is the set of all influential bones $\mathcal{I}=\{b_i|w_i>10^{-4}\}$, respectively. Here, we assume that bone $b_i$ is influential if $w_{i}>10^{-4}$. We found that the vertices are bound to their Top-3 nearest bones with more than 90\% of total skin weights, and the influential bones among these bones account for 90\% total influential bones. To achieve a balance between prediction accuracy and computation efficiency, we choose $K=3$. The smooth factor in the loss function is empirically set to $0.4$ according to grid search.

\subsection{Evaluation Metrics}

To evaluate the network's ability to predict skin weights, we leverage the following quantitative metrics:

\paragraph{Precision and Recall} The two metrics indicate the ability to find bones influential to vertices, which have similar meanings to those in pattern recognition. Precision is the fraction of correctly predicted influential bones in all predicted influential bones, and recall is the fraction of those in all influential bones in the ground truth.

\paragraph{L1-norm} This metric is used to measure the numerical accuracy of predicted weights, i.e., the distance between predicted skin weights and the ground truth. A character model's L1-norm is the average of L1-norms over its vertices.

\paragraph{Distance error} Besides directly measuring the similarity between the predicted results and the ground truth, we evaluate the deformation quality of models with our predicted skin weights. This metric represents the distance between deformed meshes, skinned with predicted skin weights and the ground truth. To simulate possible poses in real motions, we generate new poses by rotating 30\% of bones of a model by angles sampled independently from normal distributions $\mathcal{N}(0,25^\circ)$. For each character model, we generate 10 new poses, some of which are shown in Figure \ref{anim_err}. The generated poses may appear globally invalid, but the local deformations are meaningful and possibly seen in real motions. The distance error is the average of the Euclidean distances between the corresponding vertices of the two meshes.

\subsection{Model Accuracy}

\begin{table}[htbp]
  \caption{Evaluation of the predicted skin weights for our example characters.}
  \begin{tabular}{c|cccc}
    \toprule
    Character        & Precision  & Recall & L1-norm & Dist. Err. \\
    \midrule
    1          & 79.15\%     & 91.53\%  & 0.2715   &  0.002127\\
    2          & 64.86\%     & 83.67\%  & 0.5661   &  0.007667\\
    3          & 83.48\%     & 83.02\%  & 0.2795   &  0.004147\\
    4          & 88.91\%     & 89.17\%  & 0.1645   &  0.001231\\
    \bottomrule
  \end{tabular}
  \label{statistics_example}
\end{table}

\begin{figure}[htbp]
  \centering
  \includegraphics[width=\linewidth]{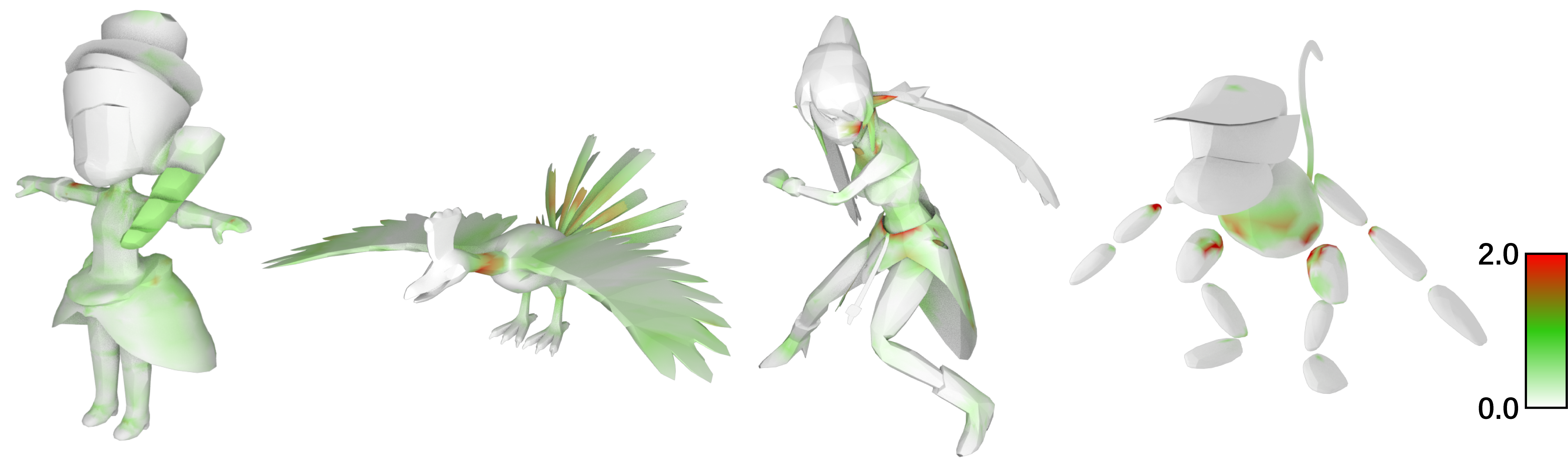}
  \caption{Per-vertex L1-norm errors on example models.}
  \label{example_l1}
\end{figure}

The overall quantitative metrics over the test set is shown in Table \ref{quantitative}. The metrics for our example characters are listed in Table \ref{statistics_example}. The results show that our network can effectively predict the skin weights for new characters with low errors in general. We then investigate the worst cases. Figure \ref{example_l1} visualizes the per-vertex L1-norm errors on example character meshes. High errors are found on the large curly hair of character 1, the neck and tail of character 2 and the belly of character 4. A further investigation shows that there is a scarcity in the training data which contains similar geometries as the ones with high predicted errors shown in Figure \ref{example_l1}, which can be improved by incorporating more data. Besides, errors tend to appear on regions near joints, such as shoulders, ankles and torsos, where vertices are influenced by multiple connected bones. Different artists tend to have different understanding of the anatomies of those areas and thus paint different skin weights on similar areas, which introduces noises to the training data.

\begin{figure}[htbp]
  \centering
  \includegraphics[width=\linewidth]{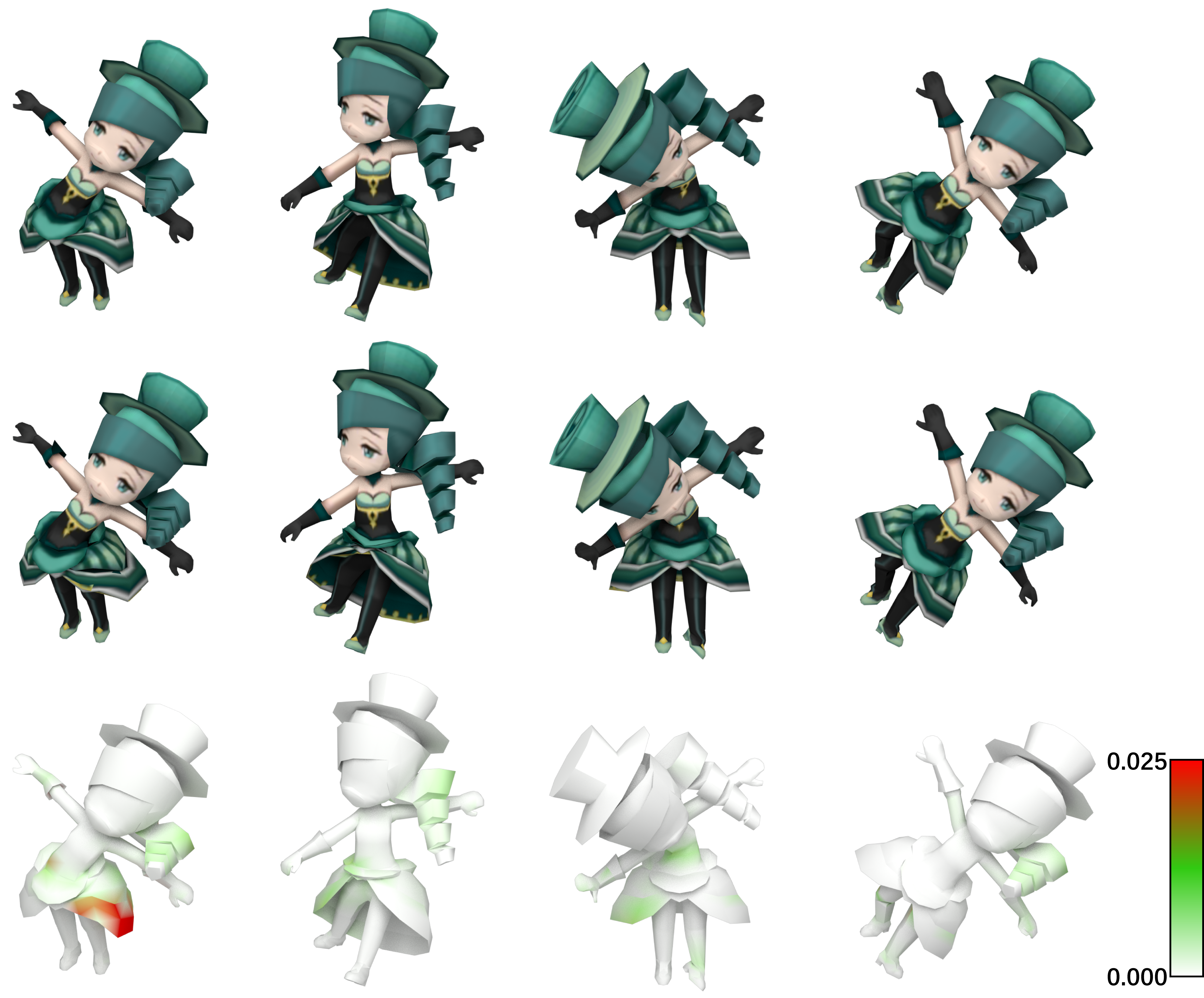}
  \caption{Randomly generated poses for a character. The meshes on the top row are deformed by the ground truth, meshes on the middle row are deformed by our predicted skin weights, and the color maps on the bottom row indicate per-vertex distance errors.}
  \label{anim_err}
\end{figure}

Figure \ref{anim_err} shows a side-by-side comparison of 4 randomly generated poses of character 1 with our predicted skin weights and the ground truth, as well as the error maps between the corresponding meshes. The largest errors are mainly on the end of the dress due to the lack of similar structures in the training set. Even with those errors, the deformation results are visually plausible and similar to those generated by the ground truth, which demonstrates that our network can predict ready-to-use skin weights for high-quality deformations.

\begin{table}[htbp]
  \caption{Comparisons with other skinning methods}
  \begin{tabular}{ccccc}
    \toprule
    Method        & Precision $\uparrow$  & Recall $\uparrow$ & L1-norm $\downarrow$ & Dist. Err. $\downarrow$ \\
    \midrule
    GeoVoxel      & 73.29\%   & 72.21\% & 0.6057 & 0.010342\\
    NeuroSkinning & 75.25\%   & 76.07\% & 0.5460 & 0.008487\\
    RigNet        & 79.94\%   & 79.02\% & 0.4075 & 0.006823\\
    Ours          & \textbf{83.04\%}    & \textbf{81.11\%}  & \textbf{0.3269}   & \textbf{0.005682}\\
    \bottomrule
  \end{tabular}
  \label{quantitative}
\end{table}

\subsection{Comparisons}

We compare our method with Geodesic Voxel binding (GVB) \cite{dionne2013geodesic}, NeuroSkinning \cite{neuroskinning} and RigNet \cite{xu2020rignet}. For GVB, we use the implementation in Maya \cite{maya} with 3 influential bones, 0.5 falloff and 128 resolution. For NeuroSkinning, since the models in the dataset have different skeletal morphologies and cannot fit into a super-skeleton, we only use their network architecture and their way in selecting the nearest bones, i.e., based on the Euclidean distance. For the distance used in RigNet, we implement the volumetric geodesic distance instead of the approximation used in their public implementation to robustly compute the distance of vertices on disconnected components and intersected triangles. We implement the above two networks using PyTorch Geometric and train and evaluate on the same train/test split as ours with the same hyperparameters (3 influential bones, 0.25 weight pruning threshold).

\begin{figure}[htbp]
  \centering
  \includegraphics[width=\linewidth]{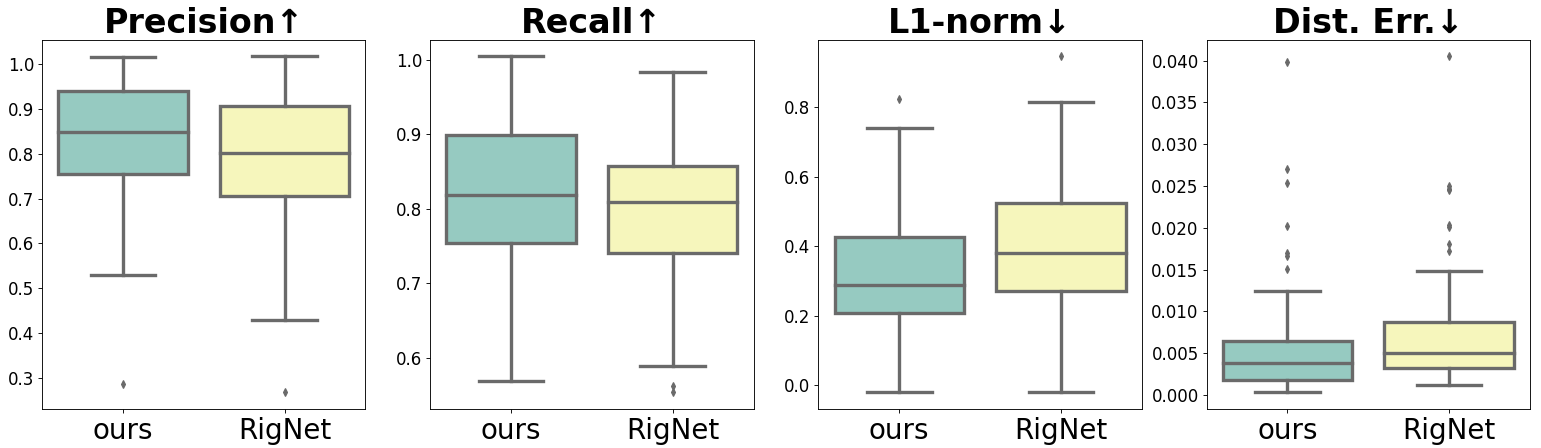}
  \caption{Statistical comparisons between our method and RigNet \cite{xu2020rignet}.}
  \label{stat}
\end{figure}

We first show the quantitative results in Table \ref{quantitative}, where the metrics are averaged over the whole test set. Our numerical results outperform the competing methods by large margins according to all quantitative metrics. We further provide statistical analysis between our method and RigNet \cite{xu2020rignet} in Figure \ref{stat}. One-way ANOVA tests are performed for all the four evaluation metrics, and they are all significant: for precision $F=4.5259, p=0.0346<0.05$, for recall $F=3.6959, p=0.0495<0.05$, for L1-norm $F=12.0506, p=0.0006<0.05$, for distance error $F=4.0373, p=0.0458<0.05$.

\begin{figure}[htbp]
  \centering
  \includegraphics[width=\linewidth]{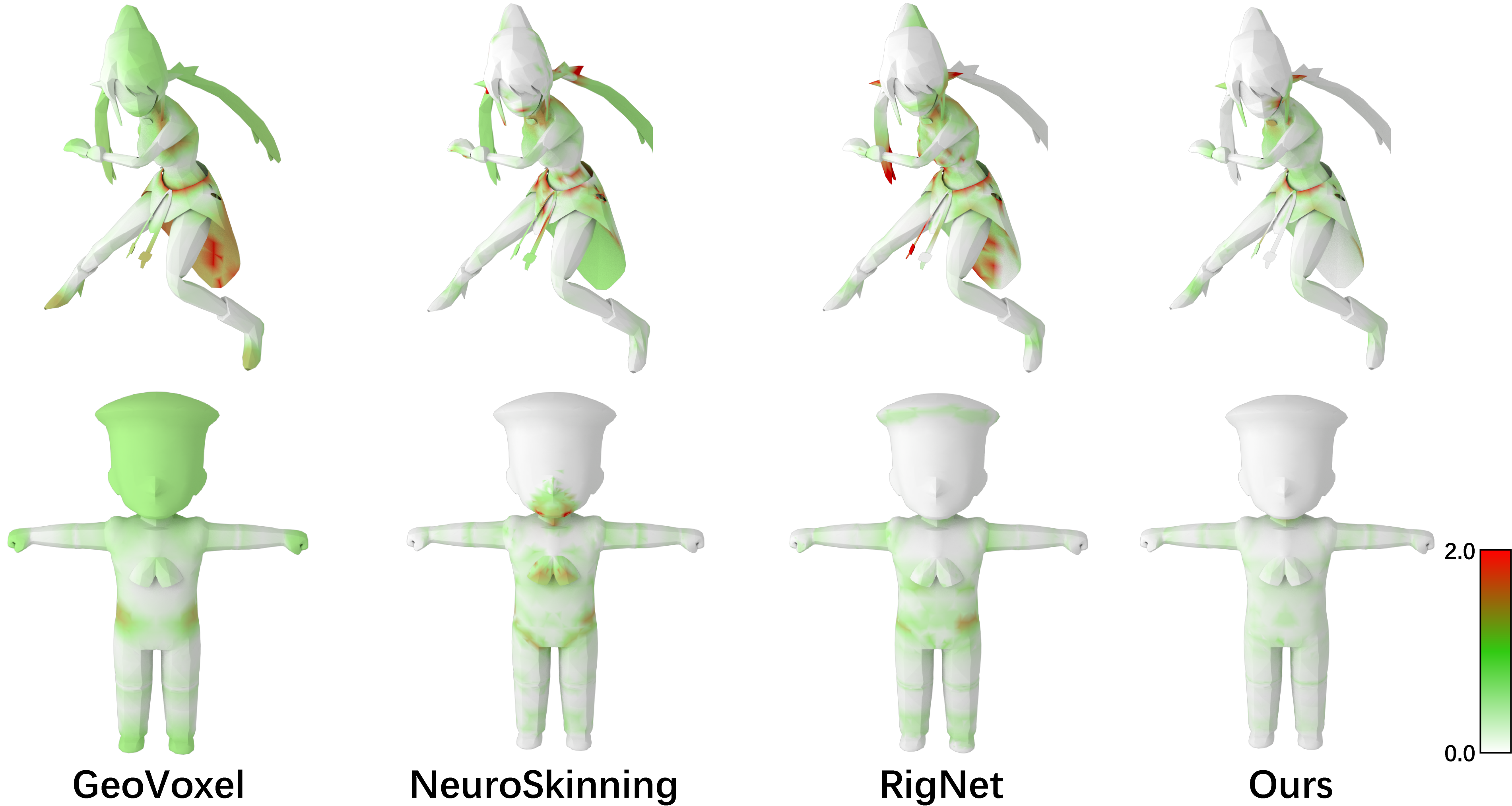}
  \caption{Comparisons of per-vertex L1-norm errors.}
  \label{L1error}
\end{figure}

Next, we qualitatively compare our method and competing methods. Figure \ref{L1error} shows a side-by-side comparison of per-vertex L1-norm errors of skin weights produced by our method and the competing methods. Our method tends to predict skin weights with lower errors. On the top example, errors tend to appear in thin parts, such as the end of the dress and the pigtails. The competing methods ignores the difference between these areas and the body parts and applies the same functions on the whole mesh. Our method extracts both bone features and vertex features and applies functions on these areas according to their bone features, hence estimates the skin weights more close to the ground truth. In the bottom example, errors tend to appear on joint areas such as shoulders, the crotch, and ankles, where vertices are influenced by multiple connected bones. Our method operates on the skeleton graphs and learns the relations between connected bones and thus produces lower errors on these areas than other methods.

\begin{figure}[htbp]
  \centering
  \includegraphics[width=\linewidth]{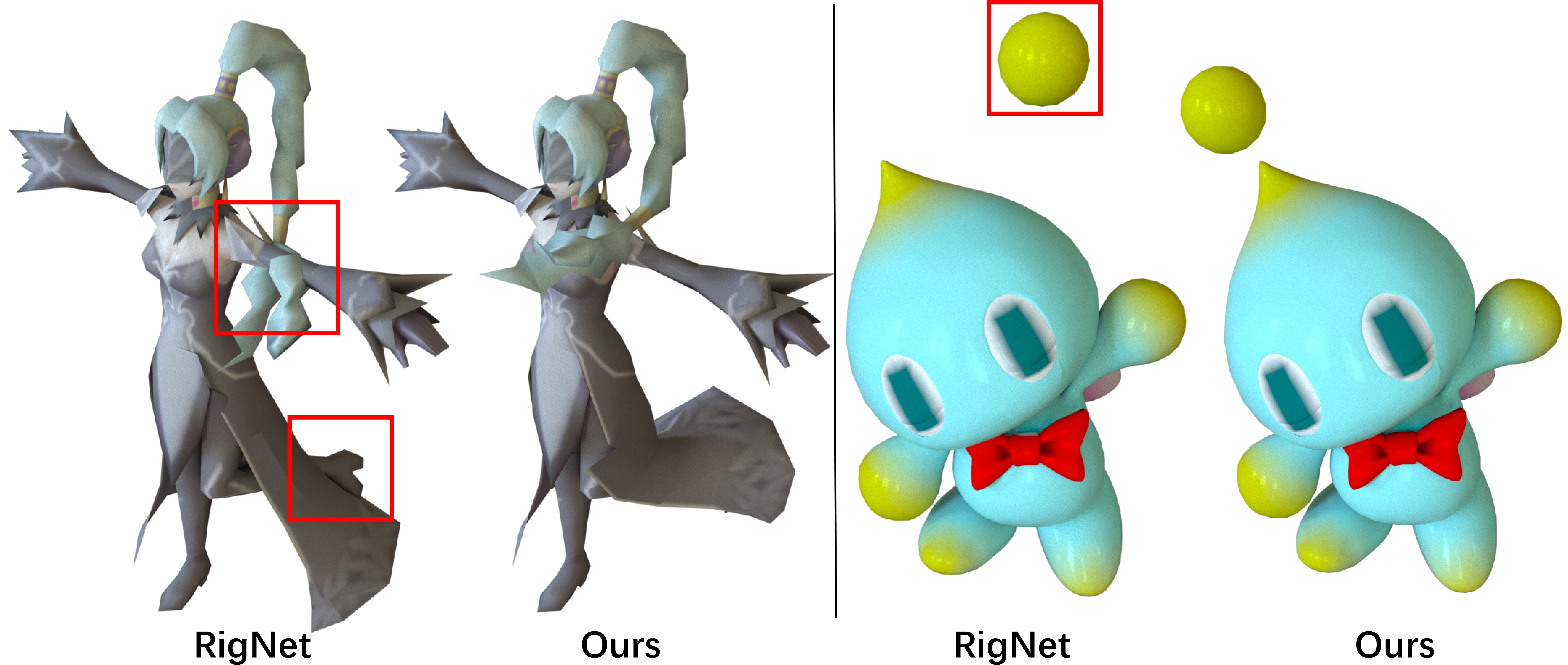}
  \caption{Comparisons of deformation results using the skin weights predicted by RigNet \cite{xu2020rignet} and our method.}
  \label{fail}
\end{figure}

We provide a closer comparison of deformation results with skin weights predicted by RigNet \cite{xu2020rignet} and our method in Figure \ref{fail}. In the left column, the ending of the braid and the bottom of the dress are controlled by out-of-body bones, RigNet fails to tackle these bones, making these areas clip with the body. The main reason is that the volumetric geodesic distance they used requires that the bones are inside the body mesh. Our method uses HollowDist, which have no assumption for bone-mesh relationships and thus is robust for out-of-body bones. In the right column, the ball is not connected with character's head but still controlled by the head bone. RigNet does not assign any bone for the ball, as it cannot calculate volumetric geodesic distances for vertices on it. Our method uses HollowDist, which handles the disconnected ball and assigns it to the head bone.

\subsection{Ablation Study}

We conduct ablation experiments to analyze the importance of each part in our network. Their quantitative results are shown in Table \ref{abalation}. First, we use the Euclidean distance as an alternative to our proposed HollowDist. The quantitative results show that the network using HollowDist significantly outperforms that using the Euclidean distance, as HollowDist captures the intrinsic properties of the meshes. Next, we remove the inter-graph convolution component, which makes the network extract the vertex and bone features separately. The quantitative result shows that this component helps the network to better extract the node features and thus obtain better prediction results. 

\begin{figure}[htbp]
  \centering
  \includegraphics[width=\linewidth]{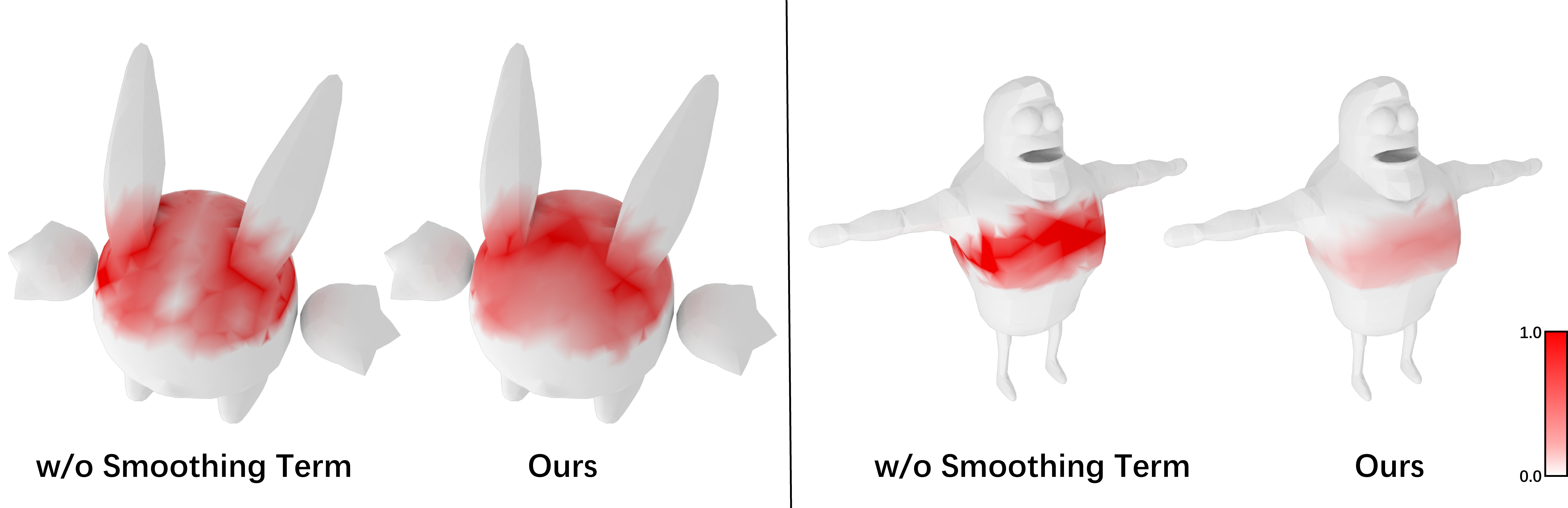}
  \caption{Comparisons of the skin weights estimated by our network with and without the smoothing term in the loss function.}
  \label{smooth_ablation}
\end{figure}

The prediction results of our network with and without the smoothing term in the loss function are shown in Figure \ref{smooth_ablation}. The smoothing term in the loss function improves the smoothness of predicted skin weights, which makes the deformations more satisfactory.

\begin{table}[htbp]
  \caption{Ablation study of our network}
  \begin{tabular}{ccccc}
    \toprule
    Method        & Precision $\uparrow$  & Recall $\uparrow$ & L1-norm $\downarrow$ & Dist. Err. $\downarrow$ \\
    \midrule
    Ours          & \textbf{82.11\%}     & \textbf{83.69\%}  & \textbf{0.3269} & 0.005682  \\
    Euclidean distance & 76.23\% & 76.52\% & 0.5253 & 0.008002 \\ 
    w/o HeterConv & 80.21\% & 80.80\% & 0.3889 & 0.006243 \\
    w/o Smoothing term & 81.40\% & 83.68\% & 0.3298 & \textbf{0.005681} \\
    \bottomrule
  \end{tabular}
  \label{abalation}
\end{table}

\section{Conclusion and Future Work}

We have presented a heterogeneous graph neural network to automatically estimate skin weights for character rigging. Our inter-graph convolution operation allows feature aggregation between heterogeneous nodes, thus our network can extract both vertex and bone features. Our method can cope with models containing multiple disjoint parts or outside bones with a new distance, HollowDist. Experimental results demonstrate that the skin weights predicted by our network can be used to produce high-quality ready-for-production animations, which greatly reduces the time needed for manual labor.

\begin{wrapfigure}{r}{6cm}
  \centering
  \includegraphics[width=0.25\textwidth]{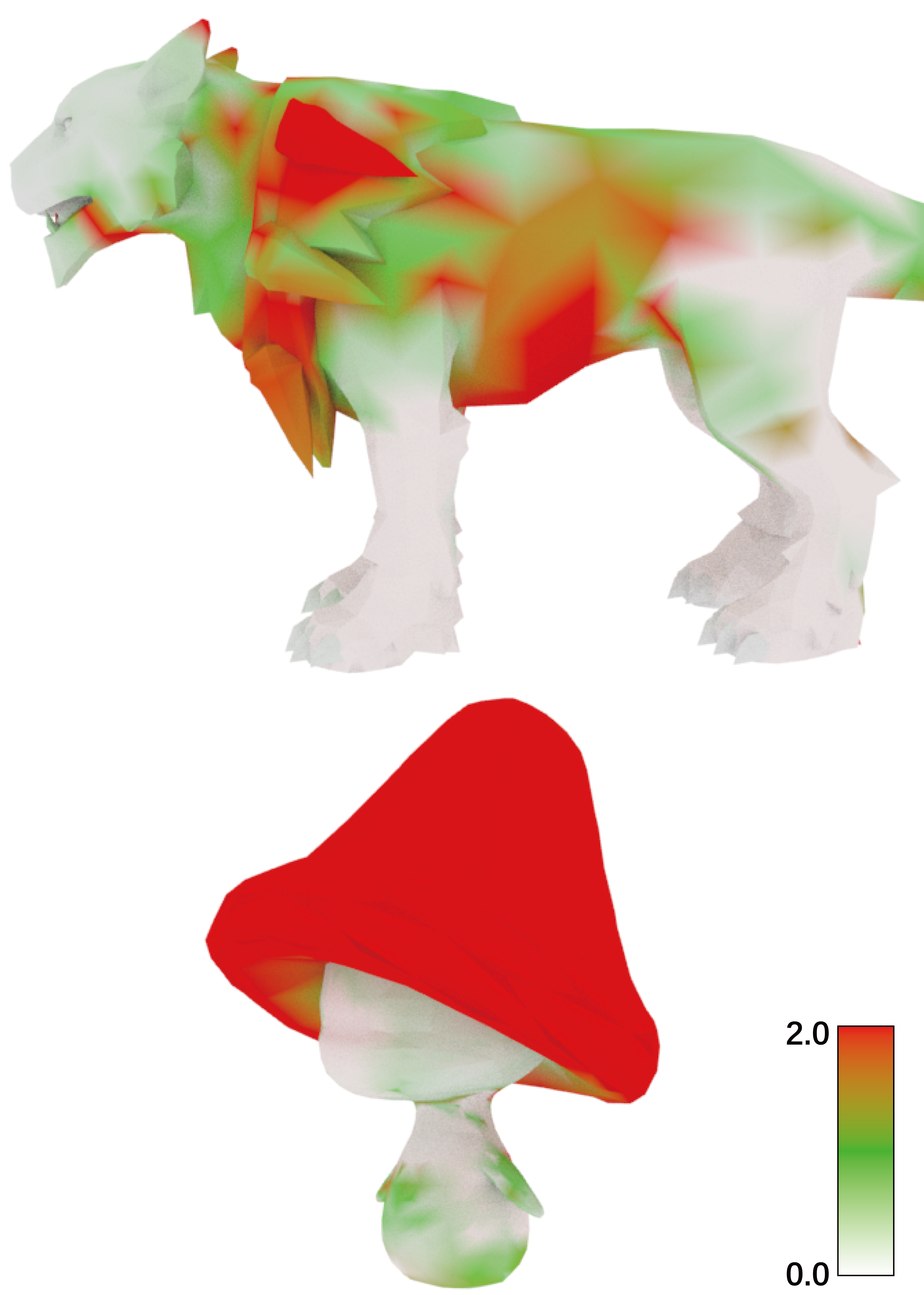}
  \caption{L1-norm errors of some failure cases.}
  \label{fail_cases}
\end{wrapfigure}

HeterSkinNet has several limitations. First, models in production are increasingly delicate and sophisticated, which may contain over 10K vertices and over 200 bones \cite{neuroskinning}. Since no such data are available, we only test our method on a dataset whose models have fewer numbers of vertices (1K-5K) and bones (10-150). Evaluating its performance on more complex models is an interesting direction for expanding its applications. Secondly, the calculation of HollowDist takes most of the time in the whole procedure (\textasciitilde 1 min). Accelerating this step would be beneficial when the number of models we process is large. Using sparse voxelization and exploring other GPU-based distances may be feasible directions. Thirdly, it assumes that a vertex is influenced by its nearest $K$ bones, where $K$ is a user-defined integer. However, the numbers of influential bones of vertices on a mesh vary according to the vertices' semantic information and the skeleton structure. It will be an interesting direction to further expand our network to predict the numbers of vertices' influential bones as well as their skin weights. Fourthly, in the vertex-to-bone convolution, we have only used the maximum, mean and variance of the vertices' features, exploring more statistics such as covariance matrix may be a valuable direction for improving the inter-graph convolution operation. Finally, for regions too far away from their control bones, such as the belly of the lion (top row of Figure \ref{fail_cases}) and the head of the mushroom (bottom row of Figure \ref{fail_cases}), our method may fail as these regions have similar HollowDists to their control bones. Exploring vertex-bone features more distinctively on these regions may boost the prediction performance.

\begin{acks}
 Xiaogang Jin was supported by the National Key R\&D Program of China (Grant No. 2017YFB1002600), the National Natural Science Foundation of China (Grant Nos. 61972344, 62036010, 61732015), the Ningbo Major Special Projects of the “Science and Technology Innovation 2025” (Grant No. 2020Z007), and the Key Research and Development Program of Zhejiang Province (Grant No. 2020C03096).
  
\end{acks}

\bibliographystyle{ACM-Reference-Format}
\bibliography{sample-base}

\end{document}